\documentclass[referee]{aa}
\usepackage{graphicx}
\usepackage{aas_macros}
\usepackage{natbib}
\usepackage{amsmath}
\usepackage{amssymb}

\bibpunct{(}{)}{;}{a}{}{,}

\begin{document}

\title{Hubble parameter reconstruction from a principal component analysis: minimizing the bias}

\author{E. E. O. Ishida\inst{1,2}  \and  R. S. de Souza\inst{1,2} }
\offprints{Emille E. O. Ishida \email{emille.ishida@ipmu.jp}}

\institute{$^1$ IAG, Universidade de S\~{a}o Paulo, Rua do Mat\~{a}o 1226, Cidade
Universit\'{a}ria, \\ CEP 05508-900, S\~{a}o Paulo, SP, Brazil\\
$^2$ Institute for the Physics and Mathematics of the Universe, University of Tokyo, Kashiwa, Chiba 277-8568, Japan}

 \date{Accepted -- Received}

\date{Released Xxxxx XX}

\authorrunning{Ishida \& de Souza}
   \titlerunning{Hubble parameter reconstruction}

\abstract
{}
{A model-independent reconstruction of the cosmic expansion rate is essential to a robust analysis of cosmological observations. Our goal is to demonstrate that current data are able to provide reasonable constraints on the behavior of the \emph{Hubble} parameter with redshift, independently of any cosmological model or underlying gravity theory.}
{Using type Ia supernova data, we show that it is possible to analytically calculate the Fisher matrix components in a \emph{Hubble} parameter analysis without assumptions about the energy content of the Universe. We used a principal component analysis  to reconstruct the \emph{Hubble} parameter as a linear combination of the Fisher matrix eigenvectors (principal components). To suppress the bias introduced by the high redshift behavior of the components, we considered the value of the \emph{Hubble} parameter at high redshift as a  free parameter. We first tested our procedure using a mock sample of type Ia supernova observations, we then applied it to the real data compiled by the Sloan Digital Sky Survey (SDSS) group.}{ In the mock sample analysis, we demonstrate that it is possible to drastically suppress the bias introduced by the high redshift behavior of the principal components. Applying our procedure to the real data, we show that it allows us to determine the behavior of the \emph{Hubble} parameter with reasonable uncertainty, without introducing any \textit{ad-hoc} parameterizations. Beyond that, our reconstruction agrees with completely independent measurements of the \emph{Hubble} parameter obtained from red-envelope galaxies.}
{}

\maketitle
\section{Introduction}

At the end of the 20th century, observations of type Ia
supernovae (SNIa) revealed that  the Universe expansion is accelerating
 \citep{riess98,perlmutter99}.  Since these publications, several efforts have been made to explain these observations (\citet{cunha09, frieman08, linder08, linder05, samsing10, freaza02, ishida05,ishida08} and references therein). In a standard analysis, dark-energy models are characterized by a small set of parameters. These are placed into the cosmic expansion rate by means of the Friedman equations, in substitution for the conventional cosmological-constant term.   This approach assumes a specific dependence of the dark-energy equation of state ($w$) on redshift and provides some insight into the probable values of the parameters involved. However, the results remain restricted to that particular parametrization.
An interesting question to attempt to answer is what can be inferred about the cosmic expansion rate from observations without any reference to a specific model for the energy content of the Universe?

To perform an independent analysis, we used
principal component analysis (PCA). In simple terms, PCA identifies the directions of data points clustering in the phase space defined by the parameters of a given
model. Consequently, it allows a dimensionality  reduction with as minimum an information loss as possible \citep{tegmark97}. The importance of a model-independent reconstruction of the cosmic expansion rate has already been investigated in the literature \citep{Huterer99,Huterer00,Tegmark02,Wang05,mignone08}. In this context, PCA has been used to reconstruct the dark-energy equation of state \citep{huterer03,Crittenden09,simpson06} and the deceleration parameter \citep{shapiro06} as a function of redshift. The use of PCA was also proposed in the interpretation of future experiments results by \citet{albretch09}. In the face of growing interest in the application of PCA to cosmology, \citet{kitching09} recall that some care must be taken in choosing the basic expansion functions and the interpretation assigned to the components.

The main goal of this work is to apply PCA to reconstruct directly the \emph{Hubble} parameter redshift dependence without any reference to a specific cosmological model. In this context, the eigenvectors and eigenvalues of the Fisher matrix form a new basis in which the \emph{Hubble} parameter is expanded. For the first time,  we show that it is possible to derive analytical expressions for the Fisher matrix if we focus on the \emph{Hubble} parameter ($H(z)$) as a sum of step functions. The reader should realize throughout this work that our procedure is mostly driven by the data, although there is a weak dependence of the components on our starting choices of parameter values. In other words, the functional form of each eigenvector is not of primary importance, we are more interested in how they are linearly combined. This approach allows us to avoid many interpretation problems pointed out by \citet{kitching09}. Our only assumption is that the Universe is spatially  homogeneous and isotropic and can be described by   Friedmann-Robertson-Walker (FRW) metric.

The paper is organized as follows. In section \ref{sec:PCA}, we briefly review  our knowledge of PCA and demonstrate how it can be  applied to a \emph{Hubble} parameter analysis using type Ia supernova observations. Section \ref{sec:application} shows the results obtained with a simulated supernova data set, following the standard procedure for dealing with the linear combination coefficients. We demonstrate that the quality of our results derived from the simulated data are greatly improved if we consider the \emph{Hubble} parameter  value in the upper redshift bound as a free parameter. We apply the same procedure to real type Ia supernova data compiled by the Sloan Digital Sky Survey team \citep{kessler09}. The results are shown in section \ref{sec:current_data}. Finally, in section \ref{sec:conclusions}, we present our conclusions.

\section{Principal component analysis}
\label{sec:PCA}

\subsection{The Fisher matrix}
\label{subsec:FM}

The procedure used to find the principal components (PCs) begins with the definition of the
Fisher information matrix  (\textsf{\textbf{F}}).  Owing to its relation to the covariance matrix (\textbf{\textsf{F}}=\textbf{\textsf{C}}$^{-1}$), it can be shown that the PCs and their associated uncertainties are related to the eigenvectors and eigenvalues of the Fisher matrix, respectively.

We consider that our data set is formed
by $N$  independent observations, each one characterized by a Gaussian
probability density function,
$f_{i}(x_i,\sigma_{i};\pmb{\beta})$. In our notation, $x_i$ represents
the $i-th$ measurement, $\sigma_{i}$ the uncertainty associated with it, and
$\pmb{\beta}$ is the vector of parameters of our theoretical model. In other words, we investigate a specific quantity, $x$, which can be written as a function of the parameters $\beta_i$, ($x(\pmb{\beta})$). In this context, the
likelihood function is given by $L=\prod_{i=1}^{N}f_i$
and the Fisher matrix is defined as
\begin{equation}
F_{kl}\equiv\left\langle-\frac{\partial^{2}\ln{L(\pmb{\beta})}}
{\partial\beta_{k}\partial\beta_{l}}\right\rangle.\label{eq:FMdefinicao}
\end{equation}
The brackets in equation (\ref{eq:FMdefinicao}) represents the
expectation value.

We can write \textbf{\textsf{F}}=\textbf{\textsf{D}}$^T \pmb{\Lambda}$ \textbf{\textsf{D}}, where the rows of the decorrelation matrix (\textsf{\textbf{D}})  are the eigenvectors ($\pmb{e_i}$) of \textbf{\textsf{F}},  and $\pmb{\Lambda}$ is a diagonal matrix whose non-zero elements are the eigenvalues ($\lambda_i$) of \textsf{\textbf{F}}. Choosing \textbf{\textsf{D}} to be orthogonal, with \textit{det}(\textsf{\textbf{D}})=1, $\pmb{e_i}$ forms an orthonormal basis of decorrelated vectors (or modes). After finding  the eigenvectors and eigenvalues of \textbf{\textsf{F}}, we rewrite $x$ as a linear combination of $\pmb{e_i}$. Our ability to determine each coefficient  of this linear expansion ($\alpha_{i}$) is given by $\sigma_{\alpha_i}=\lambda_i^{-1/2}$. Following the standard convention, we enumerate $\pmb{e_i}$ from the larger to the smaller associated eigenvalue.

The main goal of PCA is the dimensionality  reduction  of our initial parameter space. This arises in the number of PCs we use to rewrite $x$. The most accurately determined modes (smaller $\sigma_{\alpha_i}$) correspond to directions of high data clustering in the original parameter space. As a consequence, they represent a larger part of the variance present in the original data set. In the same way, the most poorly determined modes correspond to a small portion of the variance in the data, describing features that might not be important in our particular analysis. In this context, we must determine the number of PCs that will be used in the reconstruction. Our decision must be balanced between how much information we are willing to discard and the amount of uncertainty that will not compromise our results. The constraint on $x$ reconstructed with $M$ modes (where $M\leqslant N_{PC}$ and $N_{PC}$ is the total number of PCs),  is given by a simple error propagation of the uncertainties associated with each PC  \citep{huterer03}
\begin{equation}
\sigma_{PCS}^2(z)\equiv\sum_{i=1}^{N_{PC}}\left[\sigma_{\alpha_i}\pmb{e_i}(z)\right]^2\approx\sum_{i=1}^{M}\left[\sigma_{\alpha_i}\pmb{e_i}(z)\right]^2.
\label{eq:sigma_rec}
\end{equation}
From this expression, it is clear that adding one more PC adds also its associated uncertainty. At this point, we note that to calculate $F_{kl}$ we must choose numerical values for each parameter $\beta_i$. This corresponds to specifying a base model as our starting point. As a consequence, the results provided by PCA are interpreted as deviations from this initial model. The uncertainty derived from fitting the data to this base model should also be added in quadrature to equation (\ref{eq:sigma_rec}),  to compute the total uncertainty in the final reconstruction.

The question of how many PCs should be used in the final reconstruction is far from simple, and there is no standard quantitative procedure to determine it. In many cases, the decision depends on the particular data set analyzed and our expectation towards them (for a complete review see \citet{jollife02}, chapter 6). One practical way of facing the problem is to consider how many components are inconsistent with zero in a particular reconstruction. In most cases, the coefficients $\alpha_i$ tend to decrease in modulus for higher $i$, at the same time as the uncertainties associated with them increases. In this context, we can choose the final reconstruction as the one whose coefficients are all  inconsistent with zero.

The determination of one final reconstruction is beyond the scope of this work. However,  to provide an idea of how much of the initial variance is included in our plots, we shall order them following their \textit{cumulative percentage of total variance}.

The total variance present in the data is represented well  by the sum of all $\lambda_i$, and a reconstruction with the first $M$ PCs encloses a percentage of this value ($t_M$), given by
\begin{equation}
t_M=100 \frac{\sum_{i=1}^M \lambda_i}{\sum_{j=1}^{N_{PC}}\lambda_j}.\label{eq:tM}
\end{equation}
As a consequence, the question of how many PCs turns into a matter of what percentage of total variance we are willing to enclose.

\subsection{Investigating the Hubble parameter from SNIa observations}

From now on, we consider the distance modulus, $\mu$, provided by type Ia supernova observations as our observed quantity ($x_i=\mu_i$). In a very simple approach, if we consider a flat, homogeneous and isotropic Universe, described by the FRW metric, the distance modulus relates to cosmology according to
\begin{eqnarray}
\mu(z)&=&5 \log_{10}\left[d_L(z)\right] + \mu_0, \label{eq:dist_mod_ap}\\
d_{L}(z)&\equiv&(1+z)\int_{0}^{z}\frac{du}{H(u)}, \label{eq:dist_lum_ap}
\end{eqnarray}
where  $\mu_{0}$ is called \textit{intercept}, $d_L(z)$ is the luminosity distance, and $H(z)$ is the \emph{Hubble} parameter.   We  use $H_0=72$ $\rm km~ s^{-1} Mpc^{-1}$ as the current value of the \emph{Hubble} parameter \citep{komatsu09}.

To make $H(z)$ as general as possible, we write it as a sum of step functions
\begin{equation}
H(z;\pmb{\beta})=\sum_{i=1}^{N_{bin}}\beta_{i}c_{i}(z), \label{eq:expanso em H}
\end{equation}
where
\begin{displaymath}
c_{i}(z)=\left\{
\begin{array}{ll}
1 & \qquad \textrm{if}\qquad (i-1)\Delta z < z \leq i \Delta z \\
0 &\qquad \textrm{otherwise}
\end{array}\right.,
\end{displaymath}
$\beta_{i}$ are constants, $\pmb{\beta}$ is the vector formed by all $\beta_i$, $N_{bin}$ is the number of redshift bins and $\Delta z$ is the
width of each bin. This approach was proposed by \citet{shapiro06} in the context of deceleration parameter analysis. Although, when it is used for the \emph{Hubble} parameter,  the Fisher matrix  calculations are simplified  and we  still get pretty general results. Given the definition above, $\beta_i$ are now the
parameters of our theory. Physically, they represent the value of the
\emph{Hubble} parameter in each redshift bin. We can obviously express any
functional form using this prescription, with higher resolution for a larger number of bins.

In this context, we are able to  obtain analytical expressions for the luminosity distance
\begin{equation}
d_L(z,\pmb{\beta})=(1+z)\left[\Delta z \sum_{i=0}^{L(z)}\frac{1}{\beta_{i+1}}+\frac{z- L(z) \Delta z}{\beta_{L(z)+1}}\right],
\end{equation}
and its derivatives,
\begin{eqnarray}
\frac{\partial d_{L}(z,\pmb{\beta})}{\partial \beta_k}&=&-(1+z)\left[\Delta z\sum_{i=0}^{L(z)}\frac{\delta_{i+1,k}}{\beta_{i+1}^2}+\right.\nonumber\\
 & &\left.+\frac{\delta_{L(z)+1,k}\left(z-L(z)\Delta z\right)}{\beta_{L(z)+1}^2}\right],
\end{eqnarray}
where $L(z)$ corresponds to the integer part of $z/\Delta z$. From equations  (\ref{eq:FMdefinicao}), (\ref{eq:dist_mod_ap}), and (\ref{eq:dist_lum_ap}), we can calculate the Fisher matrix components as
\begin{eqnarray}
F_{kl}&=&\frac{25}{(\ln10)^2}\left[\sum_{i=1}^{N}\frac{1}{\left(\sigma_{data_i}d_L(z_i;\pmb{\beta})\right)^2}\times\right. \nonumber\\
& & \left.\times \frac{\partial d_L(z_i,\pmb{\beta})}{\partial \beta_k}\frac{\partial d_L(z_i,\pmb{\beta})}{\partial \beta_l}\right].
\end{eqnarray}

The \emph{Hubble} parameter may now be reconstructed as the sum of $H_{base}$ and a linear
combination of the new uncorrelated variables represented by the
eigenvectors of the Fisher matrix. Mathematically,
\begin{equation}
H_{rec}(z;\pmb{\alpha})= H_{base}(z)+\sum_{i=1}^{M}\alpha_{i}\pmb{e_{i}}(z),
\label{eq:reconstrucao H}
\end{equation}
where $\alpha_i$ are constants and $\pmb{\alpha}$ is the vector formed by all the $\alpha_i$. Using equation (\ref{eq:reconstrucao H}) in equations (\ref{eq:dist_mod_ap}) and (\ref{eq:dist_lum_ap}), we can write the reconstructed distance modulus. The data set is then used to find values for the parameters $\alpha_i$ that minimize the expression
\begin{equation}
\chi^2_1(\pmb{\alpha})=\sum_{i=1}^{N}\frac{(\mu_{i}-\mu_{rec}(z_{i};\pmb{\alpha}))^2}{2\sigma_{i}^{2}}.
\label{eq:chi2alpha}
\end{equation}
This minimization procedure will also generate an uncertainty in the value of parameters $\alpha_i$ ($\sigma_{\alpha_i}^{min}$), which should be taken into account in the final reconstruction error budget.

\section{Application}
\label{sec:application}

\subsection{Mock sample}
\label{subsec:toy_model}

To test the expressions and procedures presented before,  we used a simulated type Ia supernova data set. We consider $34$ redshift bins of $\Delta z=0.05$ ($0\leq z\leq 1.7$), each one containing $50$ supernovae. We tested configurations with a larger number of bins, but the results are consistent for any configuration with more than $\sim 25$ redshift bins. The uncertainty in the $i-th$ bin was calculated according to the prescription proposed in \citet{kim04}
\begin{equation}
\sigma_i=\sqrt{\frac{0.15^2}{50}+\left(0.02\frac{\Delta z (i - 0.5)}{1.7}\right)^2}.\label{eq:sigma_dados}
\end{equation}
We performed 1000 simulations of a flat Universe containing a cosmological constant and dark matter, with matter density parameter $\Omega_m=0.27$ as our fiducial model.

Our main goal in using this simulation is to obtain an idea of how the procedure proposed here behaves in an almost ideal scenario. It represents a simplified version of future data, as for the \textit{Joint Dark Energy Mission (JDEM)\footnote{http://jdem.lbl.gov/}}, but it is enough to allow us to check the consistency of our procedure.

\begin{figure}
\includegraphics[width=\columnwidth]{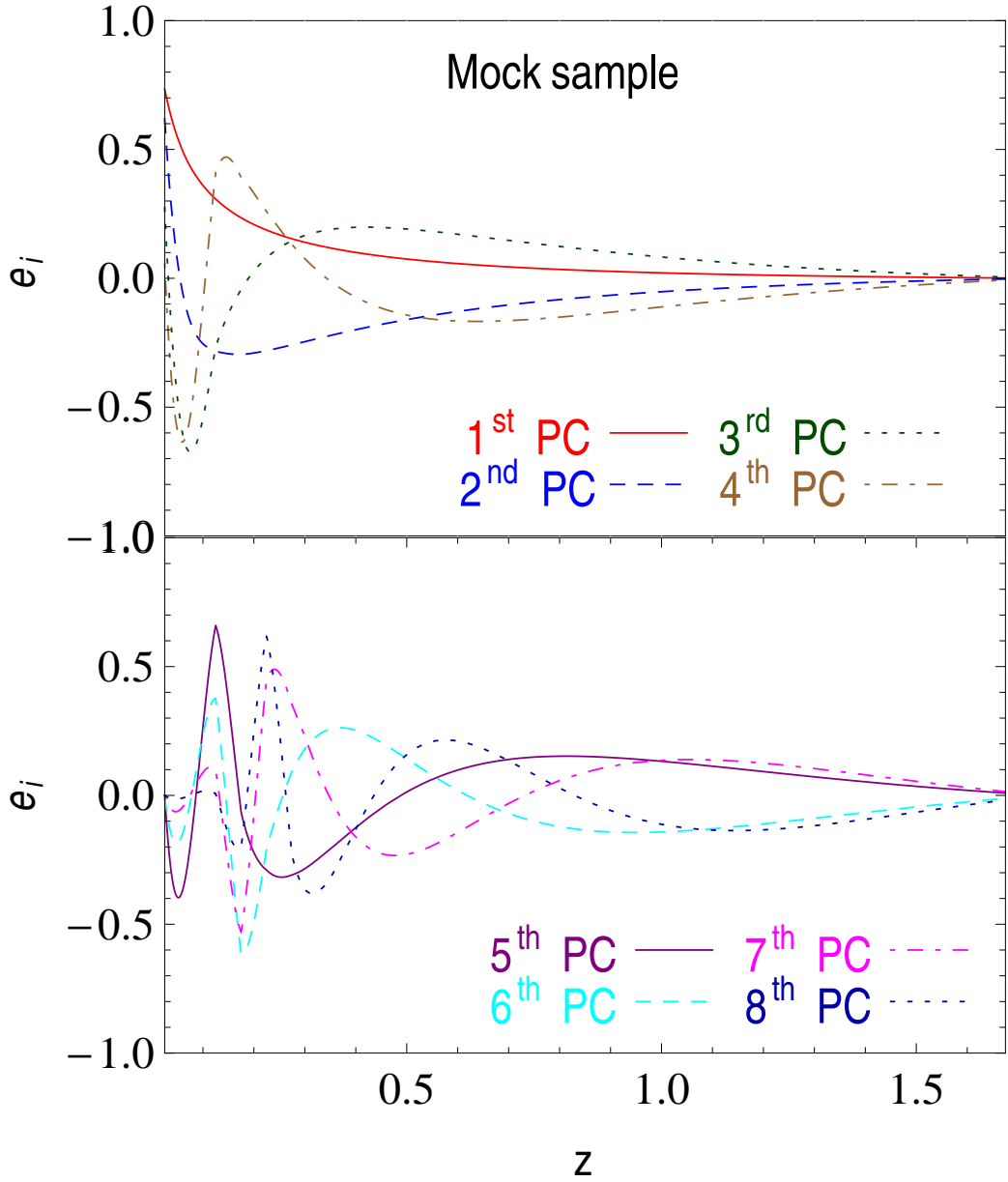}
\caption{PCs obtained from our mock sample as a function of redshift. All PCs are shown according to the convention that $\pmb{e_i}(z=0)>0$. \textbf{Top:} First (red-full),  second (blue-dashed), third (green-dotdashed), and fourth (brown-dotted) PC. \textbf{Bottom:} Fifth (purple-full), sixth (cyan-dashed), seventh (magenta-dotdashed), and eighth (dark blue-dotted) PC.}
\label{fig:PCS_toymodel}
\end{figure}

\begin{figure}
\includegraphics[width=\columnwidth]{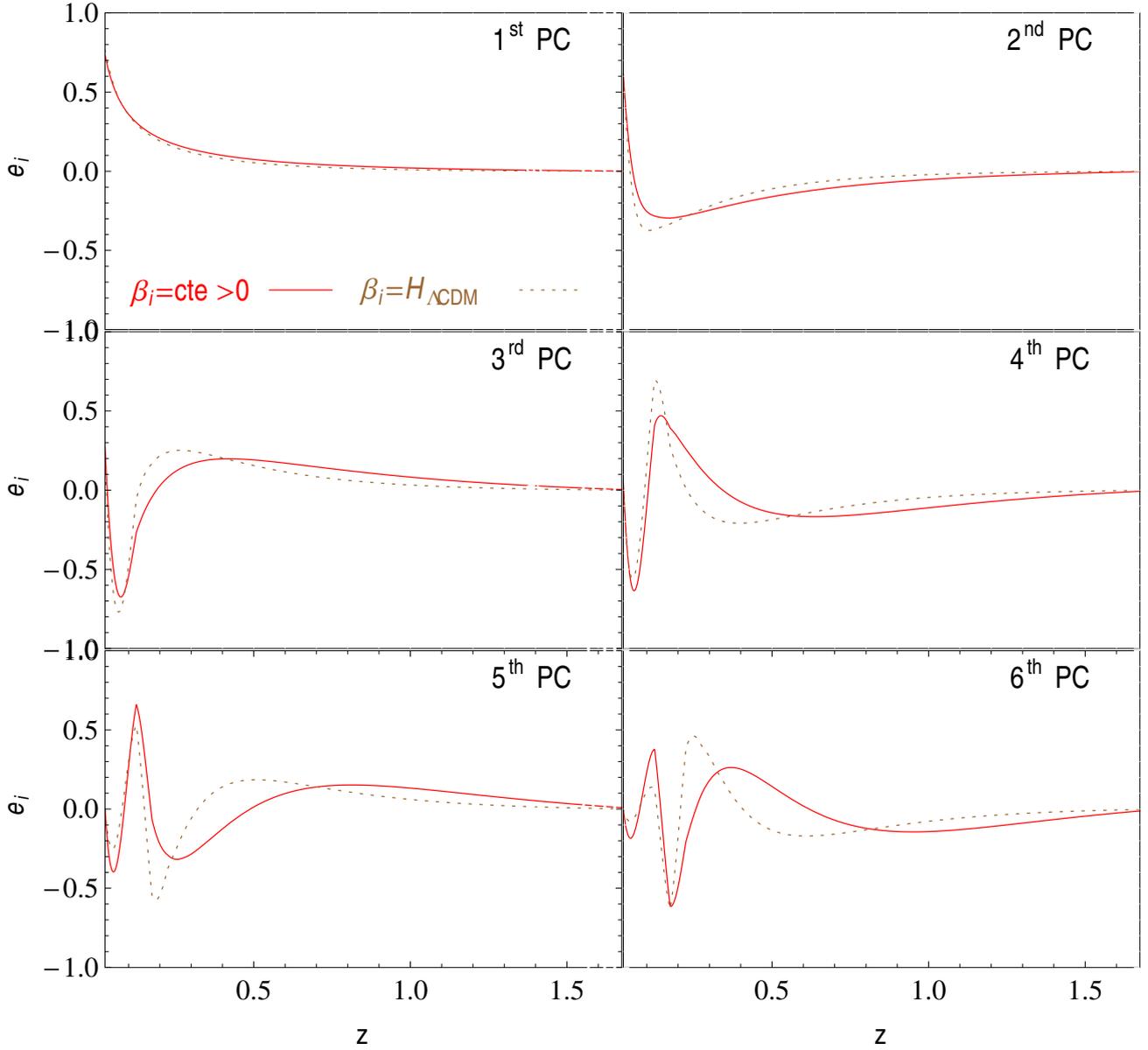}
\caption{Principal components obtained using a constant base model (full-red line) and using $\Lambda$CDM as a base model (dotted-brown line).}
\label{fig:PCS}
\end{figure}

\begin{figure}
\includegraphics[width=\columnwidth]{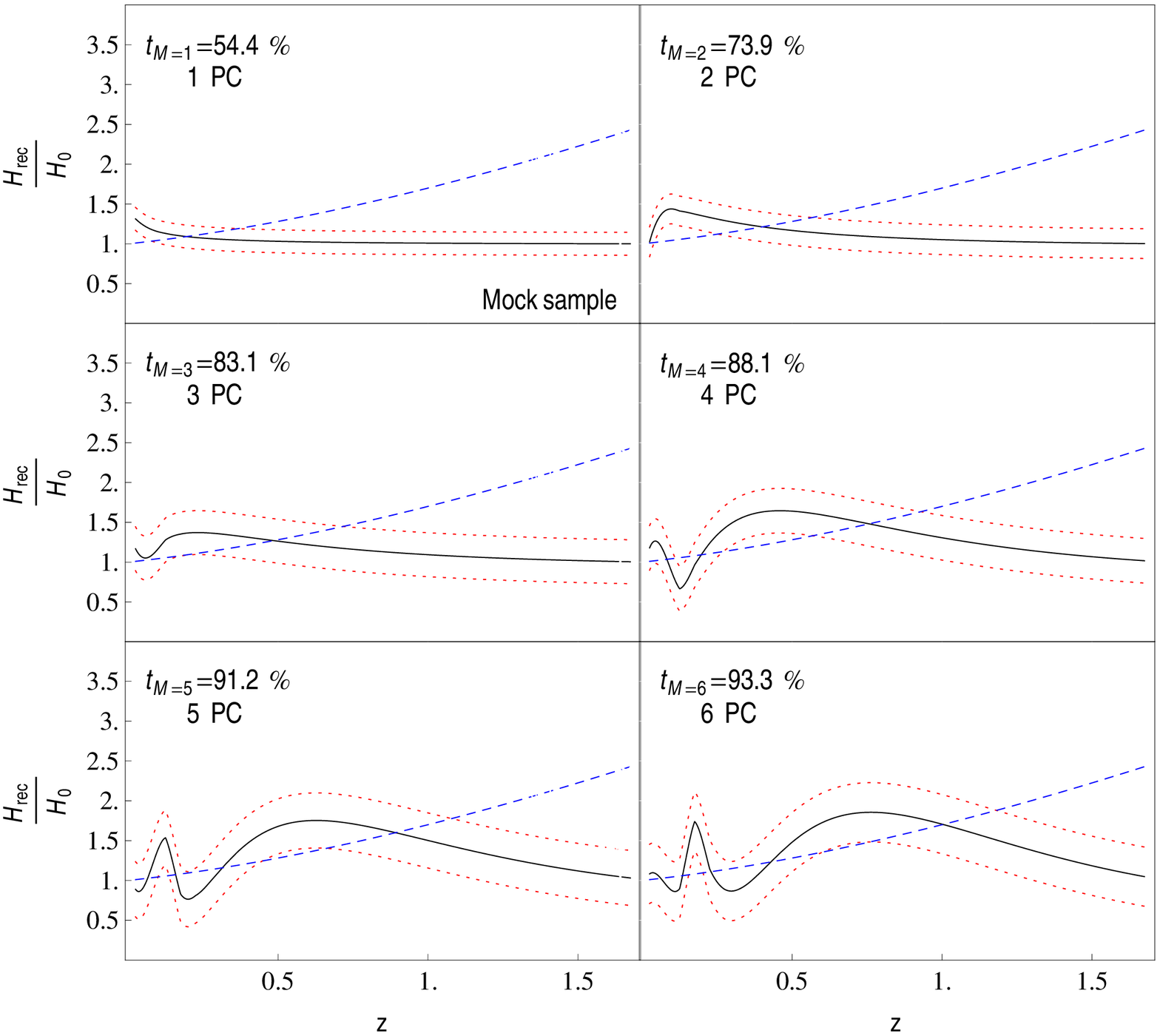}
\caption{Reconstruction of the \emph{Hubble} parameter using equations (\ref{eq:reconstrucao H}) and (\ref{eq:chi2alpha}). \emph{Hubble} parameter reconstructed with one (top-left) to six PCs (bottom-right), in units of $H_0$. The black (solid) line represents our best-fit reconstruction and the red (dotted) curves shows $2\sigma$ confidence levels. The blue (dashed) line corresponds to the the behavior of the \emph{Hubble} parameter in our fiducial model.}\label{fig:toy_rec_full}
\end{figure}

Using the equations shown in the previous section, we calculated the Fisher matrix components. We found that the modes are weakly sensitive to the choice of initial  base model  (values for the parameters $\beta_i$, hereafter $H_{base}$). However, if we use a specific cosmological model to attribute values to the parameters $\beta_i$ ($\Lambda$CDM, for example), all the results derived from this initial choice can only be analyzed in the face of that model. As our goal is to make a model-independent analysis, the best choice is to calculate the PCs based on a model where there is no evolution with redshift ($\beta_i=\textrm{cte}>0, \forall i$). The PCs will then denote deviations from a constant behavior, regardless of the value attributed to $\beta_i$. A constant \emph{Hubble} parameter is obviously an extremely unrealistic model, although, it does allow us to have a better idea of which characteristics of our results are extracted from the data and which are only  a consequence of our initial choices.

We do not currently have well constrained information about the evolution of the \emph{Hubble} parameter with redshift, but we do have independent measurements of its value today, $H_0$ (e.g. \cite{komatsu09}). Hence, we present our results in units of $H_0$ and use a base model in which $\beta_i=1, \forall i$. The resulting eigenvectors with larger eigenvalues are shown in Fig. \ref{fig:PCS_toymodel}.

The comparison between the PCs obtained from using $H_{base}=\Lambda$CDM and  $H_{base}=cte$ is shown in figure (\ref{fig:PCS}). From this plot, we can see that the difference exist, but the overall shape of the PCs are not very sensitive to the choice of $H_{base}$.

To clear illustrate the standard-procedure results of PCA reconstruction in the specific case studied here, we show in Fig. \ref{fig:toy_rec_full} reconstructions using one to six PCs with corresponding values of $t_M$.  From this plot, it is clear that our attempt to reconstruct $H(z)$ using a few PCs does not provide the expected results. We have two main problems here: the reconstruction merely oscillates around the fiducial model (blue-dashed line) and we can clearly see that there is a bias dominating the high-redshift behavior. We address both problems in the next section.

\subsection{Minimizing the bias in the reconstruction}
\label{subsec:min_bias}

From Fig. \ref{fig:PCS_toymodel}, we realize that all the first eight PCs go to zero at high-redshift, which means that at these redshifts our data provides little information \footnote{This kind of behavior is also present in PCs from the dark energy equation of state \citep{huterer03} and deceleration parameter \citep{shapiro06}.}. Consequently, no matter how many PCs we use or which values we attribute to the parameters $\alpha_i$, the reconstructed function will always be biased in the direction of our previously chosen base model for high $z$ (in this work, ``high $z$" corresponds to the upper redshift bound of our data set. In our mock sample, $z_{max}=1.7$).

\begin{figure}
\includegraphics[width=\columnwidth]{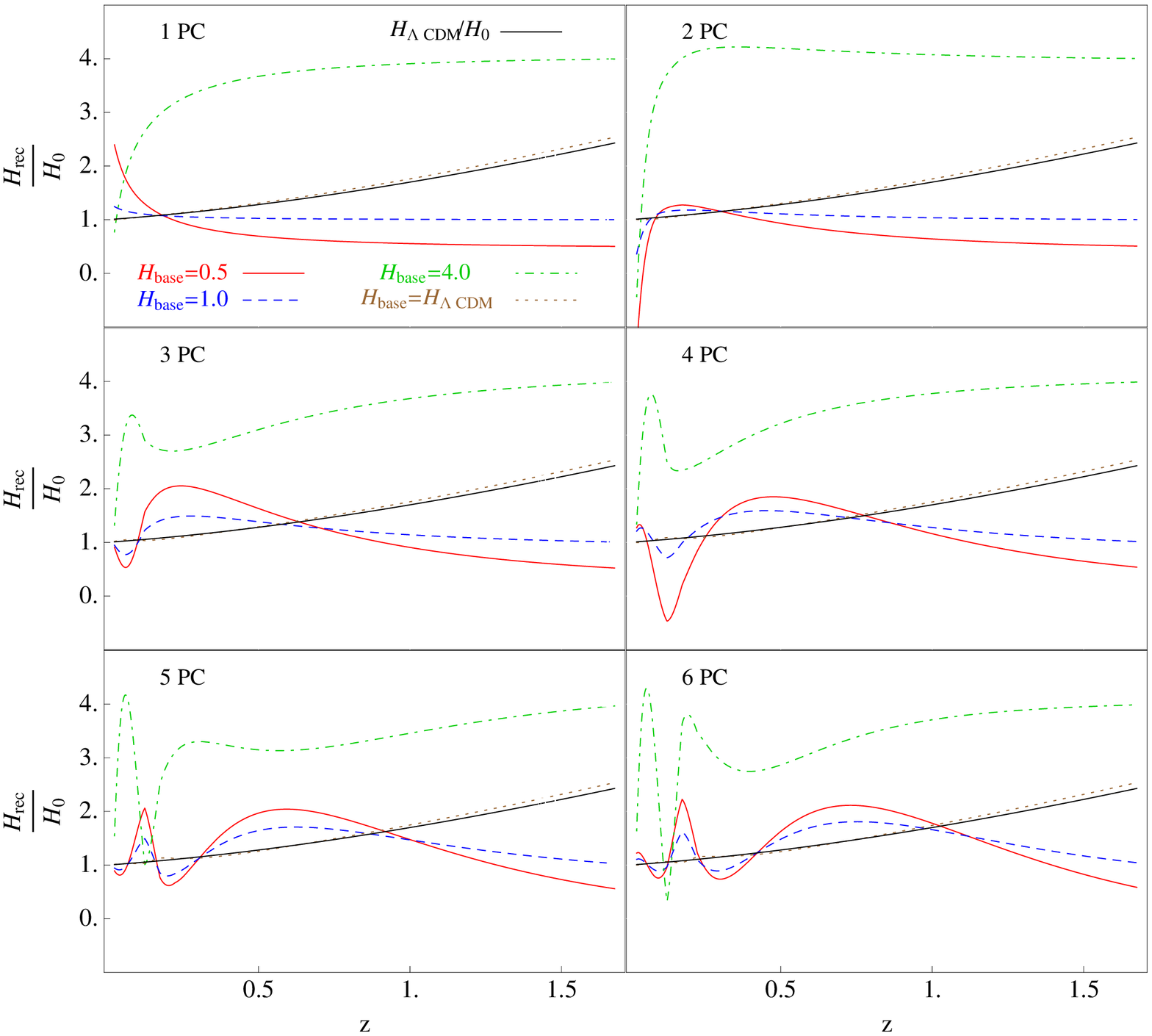}
\caption{Reconstructions using 1 (top-left) to 6 (right-bottom) PCs for different $H_{base}$, obtained from our mock sample. The black (full) line represents the fiducial model $\Lambda$CDM, the brown (dotted) line corresponds to the final reconstructions in the case $H_{base}(z)= \Lambda CDM$, the red (full) line corresponds to the case $H_{base}(z)=0.5$, the blue (dashed) line shows the reconstruction considering $H_{base}=1.0$, and the green (dot-dashed) line is the reconstruction for $H_{base}(z)=4.0.$}\label{fig:rec_diff_beta_h_fix}
\end{figure}

At this point, we must pay attention to the crucial role played by $H_{base}$ in the standard procedure described so far. Although the PCs depend weakly on our choice of $H_{base}$, the final reconstruction is extremely sensitive to the choice.  Figure (\ref{fig:rec_diff_beta_h_fix}) shows how different $H_{base}$ lead to completely different final reconstructions.

Searching the literature, we found two different approaches to dealing with this problem. We could ignore the reconstruction in the region of high redshift \citep{huterer03,shapiro06} or add a physically motivated model for $H_{base}$ in equation (\ref{eq:reconstrucao
H}), which would provide us with the value we expect to measure in the upper redshift bound \citep{tang08}. We consider that  the first alternative does not represent a good solution. The problem is not only the bias at high $z$, but also the weird behavior present in the reconstruction as a whole. Beyond that, our intention is not only to improve the fit quality, but also to make it independent of our initial choice of $H_{base}$. The second alternative would produce results in good agreement with the fiducial model (corresponding to the dotted-brown reconstruction in figure (\ref{fig:rec_diff_beta_h_fix})), in a simulated situation. Defining a  physically motivated $H_{base}$ would only, however, introduce another bias. As in reality we do not have access to the ``true" value of $H(z)$, this would require us to make a hypothesis about the energy content and dark energy model, which we are trying to avoid.

\begin{figure}
\includegraphics[width=\columnwidth]{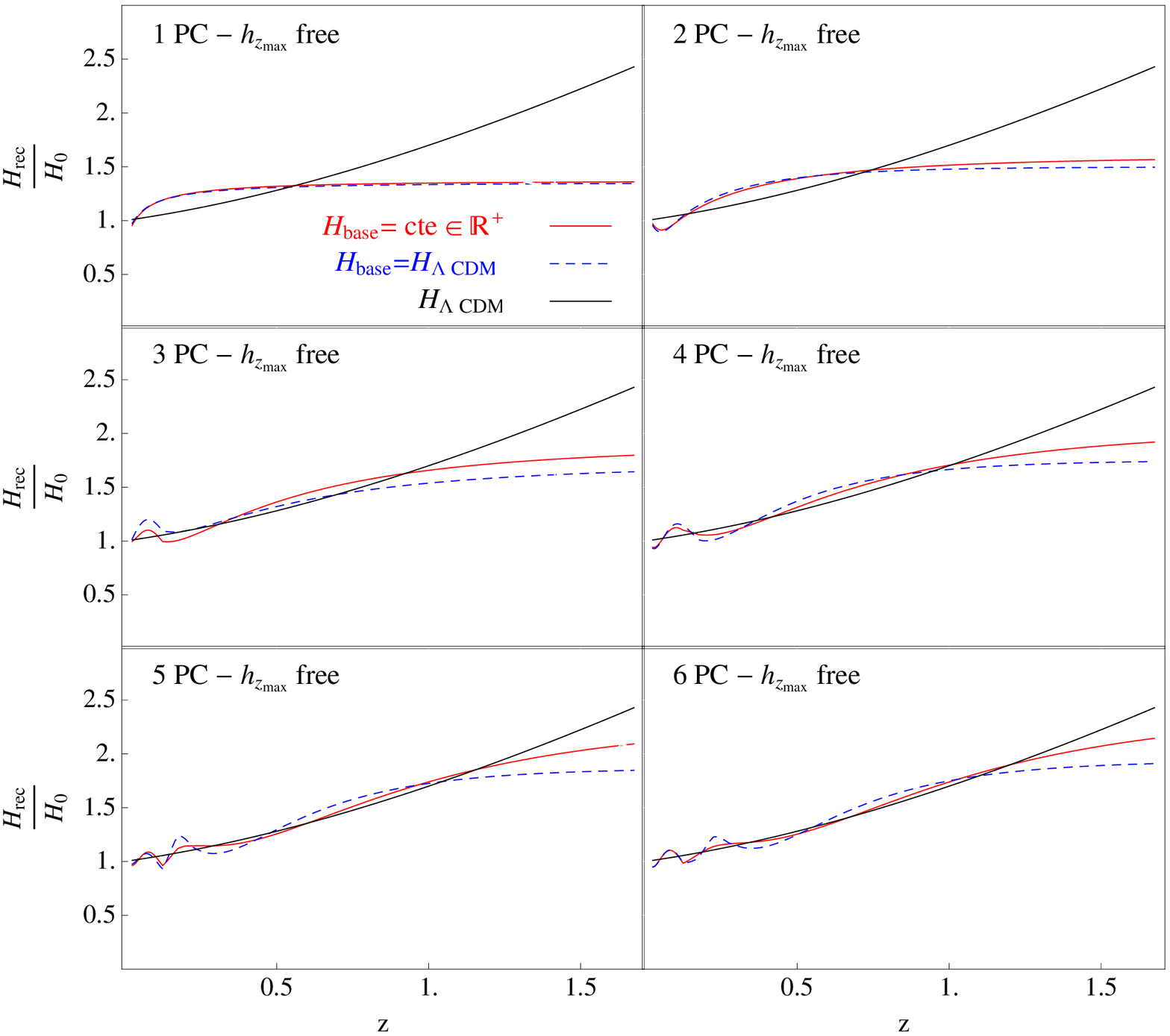}
\caption{Reconstructions using one (top-left) to six (right-bottom) PCs for different choices of $H_{base}$ and considering $h_{z_{max}}$ as a free parameter, for our mock sample. The black (full) line represents the fiducial model $\Lambda$CDM, the blue (dashed) line corresponds to the final reconstructions in the case $H_{base}(z)=H_{\Lambda CDM}$ and the red (full) line corresponds to the case $H_{base}(z)=cte>0$.}
\label{fig:rec_diff_beta_h_free}
\end{figure}

In this context, we believe that it is reasonable to consider the behavior of $H(z)$  at high $z$  as a free parameter. This means adding another parameter ($h_{z_{max}}$) to equation (\ref{eq:reconstrucao H}), which becomes
\begin{equation}
H_{rec}(z)=h_{z_{max}}+H_{base}(z)+\sum_{i=1}^N\alpha_i \pmb{e_i}(z)
\label{eq:rec_hfid}.
\end{equation}
As a consequence, the new $\chi^2$ will be given by
\begin{equation}
\chi^2(h_{z_{max}},\pmb{\alpha})=\sum_{i=1}^{N}\frac{(\mu_{i}-\mu_{rec}(z_{i};h_{z_{max}},\pmb{\alpha}))^2}{2\sigma_{i}^{2}},
\label{eq:chi2_hfid}
\end{equation}
and the uncertainty associated with the determination of $h_{z_{max}}$ ($\sigma_{h_{z_{max}}}$) is added in quadrature to the right hand side of equation (\ref{eq:sigma_rec}), leading to
\begin{equation}
\sigma_{rec}^2(z)\approx \sigma_{h_{z_{max}}}^2+\sum_{i=1}^{M}\left[\left(\sigma_{\alpha_i}\pmb{e_i}(z)\right)^2 + \left(\sigma_{\alpha_i}^{min}\right)^2\right]. \label{eq:sigma_rec_hfid}
\end{equation}
The effect of this choice is shown in figure (\ref{fig:rec_diff_beta_h_free}). We can see that, no matter which $H_{base}$ we use, if $h_{z_{max}}$ is consider to be a free parameter, the reconstruction at high $z$ is driven by the data.  As a consequence, we obtain good agreement for the reconstruction using a constant as well as a $\Lambda$CDM model for $H_{base}$.  Even though the PCs are not identical for non-evolving and $\Lambda$CDM models (figure (\ref{fig:PCS})), this agreement is a direct consequence of the best-fit values of the set of parameters $\{\alpha_i,h_{z_{max}}\}$ always arranging themselves to more accurately describe the information in the data.

\begin{figure}
\includegraphics[width=\columnwidth]{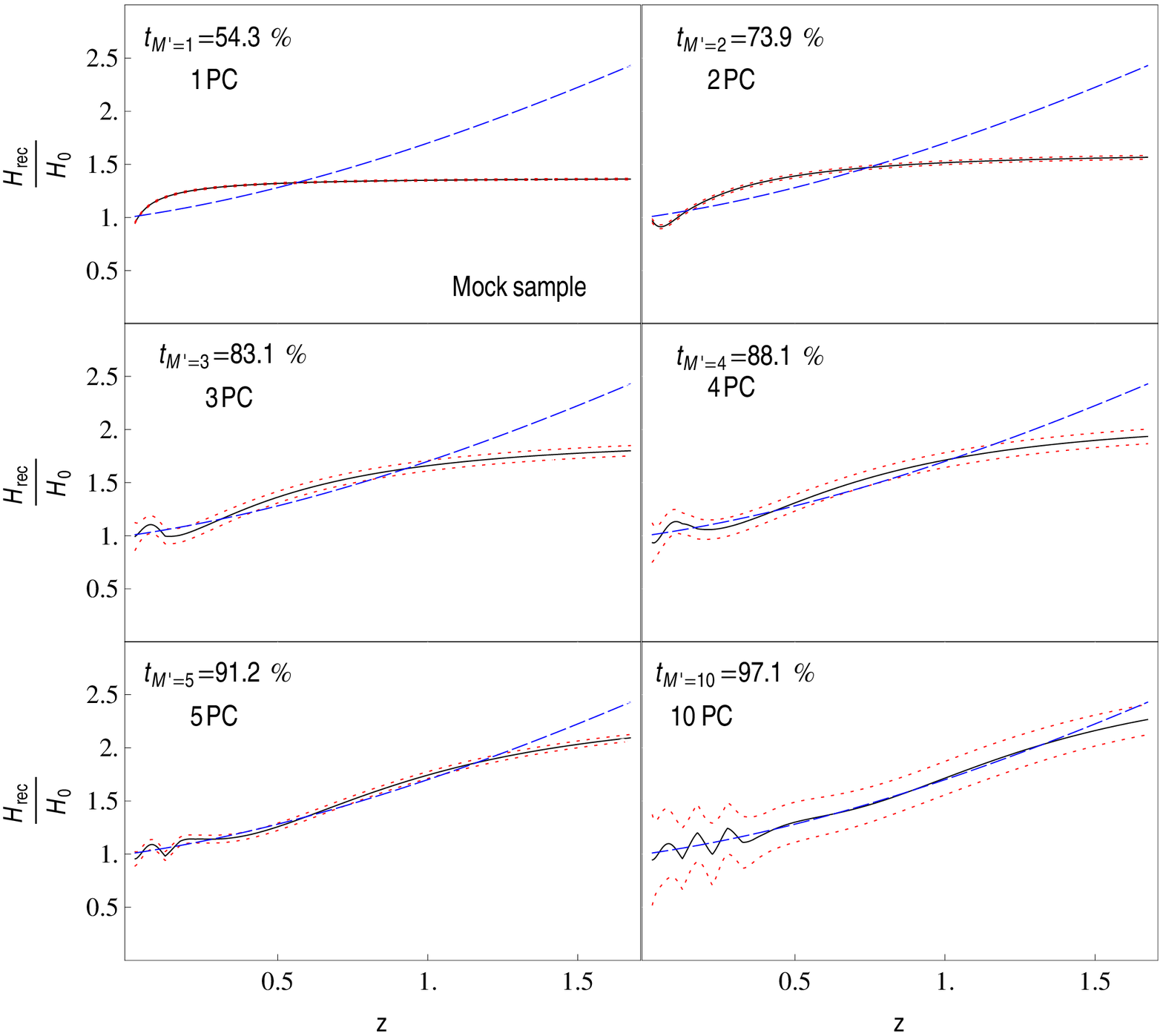}
\caption{Reconstruction of the \emph{Hubble} parameter using equations (\ref{eq:rec_hfid}) and (\ref{eq:chi2_hfid}). \emph{Hubble} parameter reconstructed with 1 (top-left) to 5 (bottom-left) and 10 (bottom-right) PCs, in units of $H_0$. The color code is the same used in figure (\ref{fig:toy_rec_full}).}\label{fig:toy_rec_hlivre}
\end{figure}

We present in Fig. \ref{fig:toy_rec_hlivre} the results of the reconstruction using 1 to 5, and 10 PCs, and corresponding values of $t_M$. Comparing these results with those in Fig. \ref{fig:toy_rec_full}, we can see a huge improvement in the agreement between the reconstructed function and the fiducial model. The reconstruction with 10 PCs encloses the fiducial model within $2\sigma$ confidence levels in the whole redshift range covered by the data. Considering that initially we had 34 parameters $\beta_i$, this represents a reduction of $\approx 70\%$ in the parameter space dimensionality.

We have so far demonstrated that PCA is an effective method for determining the \emph{Hubble} parameter behavior with redshift. It provides a considerable reduction in the initial parameter space dimensionality, without introducing any hypothesis about the energy content, cosmological model, or underlying gravity theory. The reconstruction relies on the assumption of a homogeneous and isotropic Universe, described by a FRW metric. The simulated data set used above is composed of independent data points, each one associated with a Gaussian probability density function, in a flat $\Lambda$CDM Universe.

In what follows, we apply this procedure to a real (and consequently less well behaved) data set. Our goal is to see whether, in a realistic scenario, the effectiveness of the procedure remains.

\section{Results from current SNIa data}
\label{sec:current_data}

\begin{figure}
\includegraphics[width=\columnwidth]{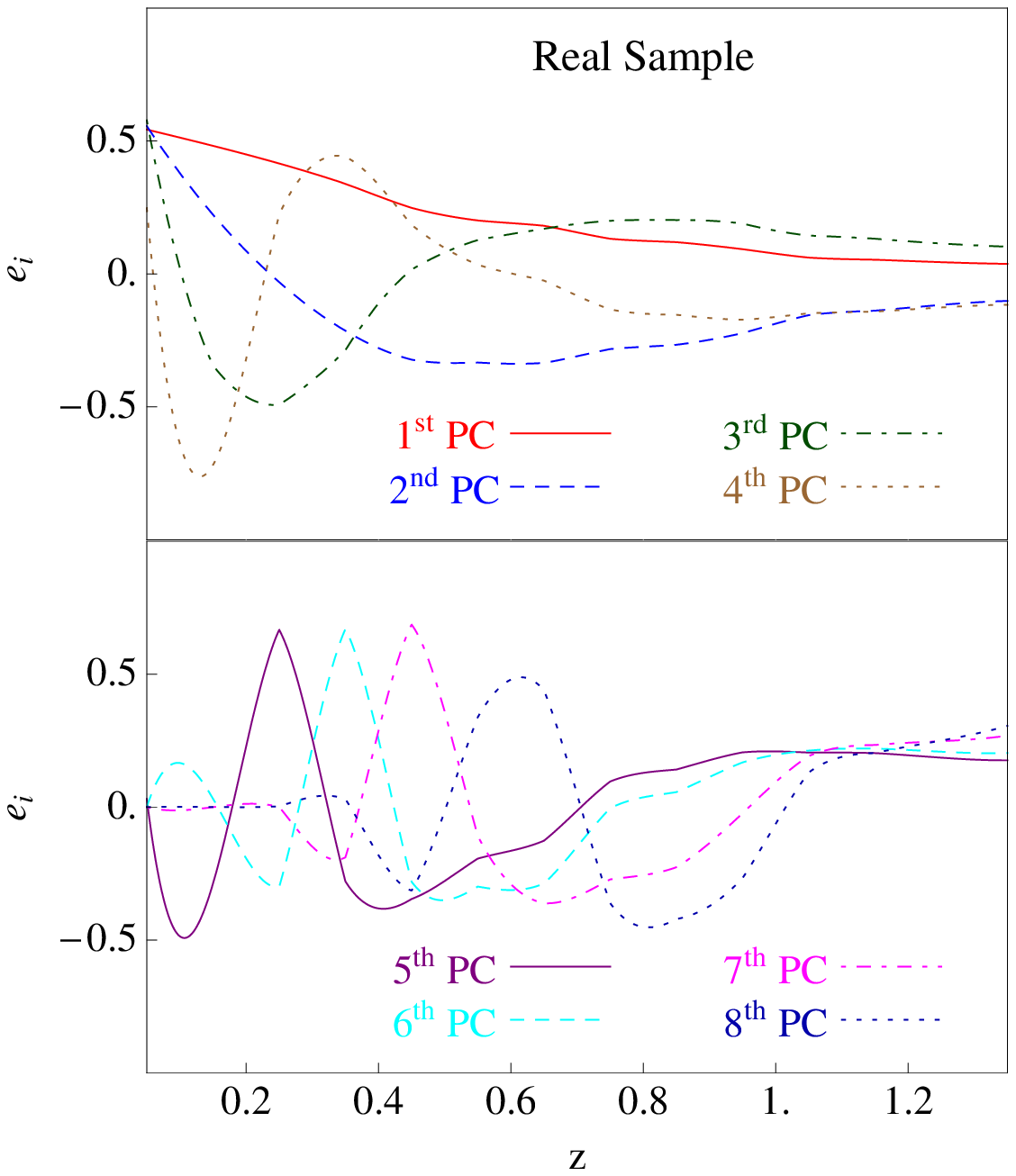}
\caption{Principal components obtained from real data sample. All PCs are plotted following the convention $\pmb{e_i}(z=0)>0$. \textbf{Top:} First (red-full), second (blue-dashed), third (green-dotdashed), and fourth (brown-dotted) PCs. \textbf{Bottom:} Fifth (purple-full), sixth (cyan-dashed), seventh (magenta-dotdashed), and eight (darker blue-dotted) PCs.} \label{fig:SDSS_PCS}
\end{figure}

\begin{figure}
\includegraphics[width=\columnwidth]{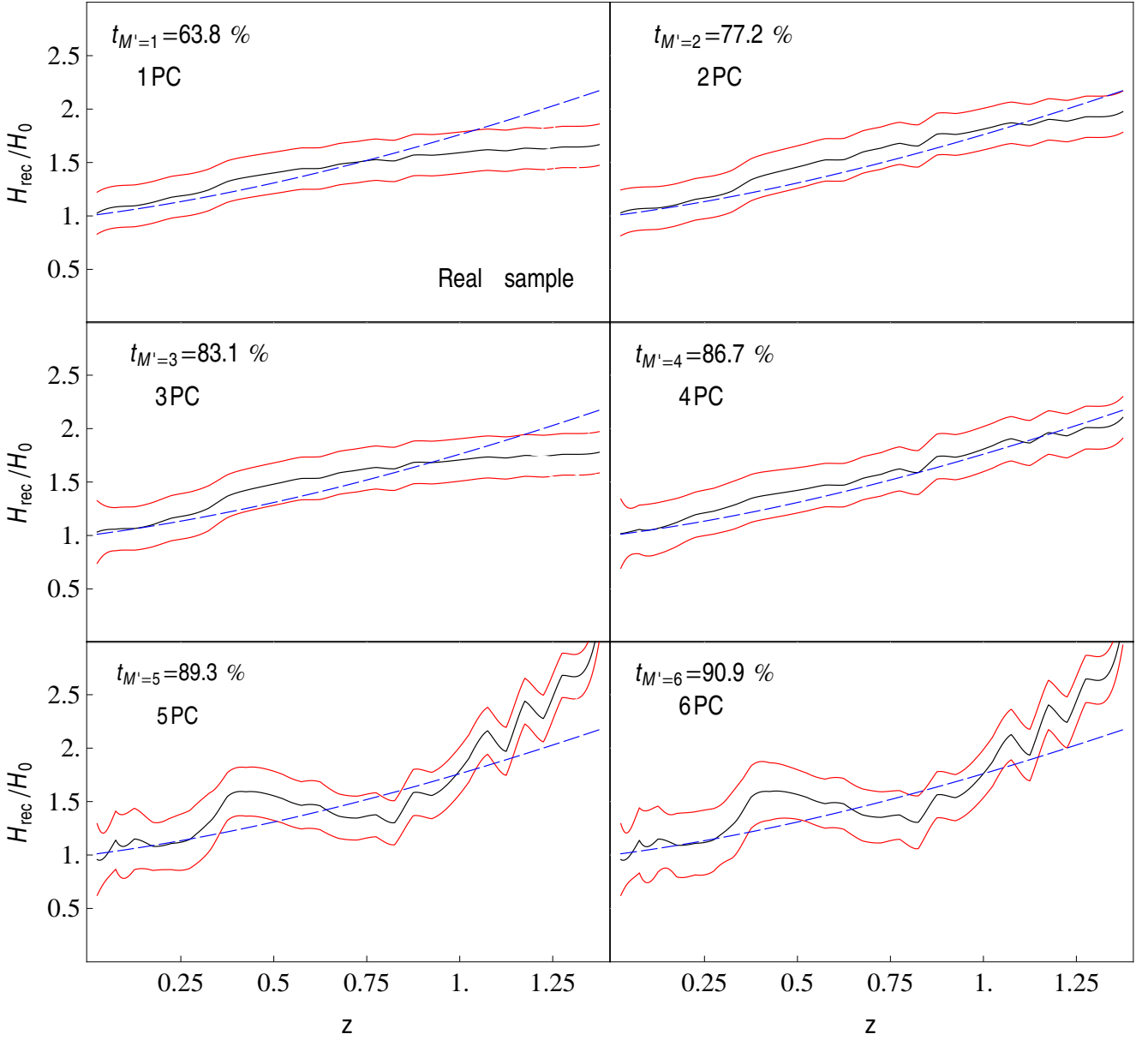}
\caption{Reconstructions obtained from linear combinations of the PCs shown in figure \ref{fig:SDSS_PCS}. Panels run from 1 (top-left) to 6 (bottom-right) PCs. The blue dashed line represents $H(z)$ in a flat XCDM cosmological model with $w=-0.76$ and $\Omega_m=0.30$ (best-fit results reported by \citet{kessler09}). Black curve corresponds to the best-fit reconstruction and red lines are $2\sigma$ confidence levels. } \label{fig:SDSS_rec6}
\end{figure}

We applied this procedure to real supernova Ia data compiled by the Sloan Digital Sky Survey (SDSS)  SN group, hereafter \textit{real sample}. This sample include measurements from the SDSS \citep{kessler09}, the ESSENCE survey \citep{miknaitis07, wood07}, the Supernova Legacy Survey (SNLS)\citep{astier06}, the \emph{Hubble} Space Telescope (HST) \citep{garnavich98, knop03, riess04, riess07}, and a compilation of nearby SN Ia measurements \citep{jha07}. The first eight PCs found from the real data set are shown in Fig. \ref{fig:SDSS_PCS}. We used 28 redshift bins of width $\Delta z=0.05$ and the 287 data points from the aforementioned data set with $z\leq 1.4$. The resulting reconstructions with one to six PCs are shown in Fig. \ref{fig:SDSS_rec6}. The blue-dashed line corresponds to the behavior of the \emph{Hubble} parameter in a flat Universe containing dark energy with an equation-of-state parameter for dark energy  $w_{dark} =-0.76$ and matter density parameter $\Omega_m=0.30$. This corresponds to the best-fit, flat cosmology found by \citet{kessler09} in the context of the \textit{Multicolor Light Curve Shape} (MLCS2k2, \citep{jha07}), hereafter fXCDM. It is shown here exclusively for comparison reasons, this model was not used in our calculations.

Comparing Figs. \ref{fig:toy_rec_hlivre} and \ref{fig:SDSS_rec6}, we realize that the confidence intervals are much larger in Fig. \ref{fig:SDSS_rec6}, as expected, because of the observational uncertainties that are present only in the real sample. Beyond that, the contours corresponding to $2\sigma$ confidence levels do not evolve in Fig. \ref{fig:SDSS_rec6} as they do in Fig. \ref{fig:toy_rec_hlivre}. This is a direct consequence of our choice of introducing $h_{z_{max}}$ as a free parameter. In the simulated case,  $\sigma_{h_{z_{max}}}$ is much smaller than the uncertainty associated with the parameters $\alpha_i$. As a consequence, the evolution of the confidence levels is dominated by the uncertainties associated with the PCs. In the real case, the opposite situation occurs. For the six cases presented in Fig. \ref{fig:SDSS_rec6}, $\sigma_{h_{z_{max}}}\gg\sigma_{\pmb{\alpha}}$, making the final uncertainty almost independent of how many PCs are used in the reconstruction. This behavior is a consequence of the low number and quality of data points at high redshift. However, it is also related to a non-null correlation between the uncertainties in determining the parameters $\alpha_i$ and $h_{z_{max}}$. To explore this method in the best case scenario, we need to ensure not only that we have high number and quality of data points at high redshift, but also that the PCs are as well determined as possible.

In comparing Figs. \ref{fig:toy_rec_hlivre} and \ref{fig:SDSS_rec6}, the reader should be aware that the blue dashed line means different things in each figure. Figure \ref{fig:toy_rec_hlivre} is a simulation, and in such a case the blue dashed line corresponds to the fiducial model used to generate our mock sample. Fig. \ref{fig:SDSS_rec6} was created using real data,  in this case the blue dashed line represents the best-fit flat $\Lambda$CDM model, as reported by \cite{kessler09}.

We can also see, in Fig. \ref{fig:SDSS_rec6}, that the reconstruction becomes more irregular when more than four PCs are used. If we take the blue dashed line as a good representation of the ``real" cosmological model, we could say that four PCs are enough to enclose the desired behavior within $2\sigma$ confidence level. For the sake of completeness, we plot reconstructions up to six PCs.

\begin{figure}
\includegraphics[width=\columnwidth]{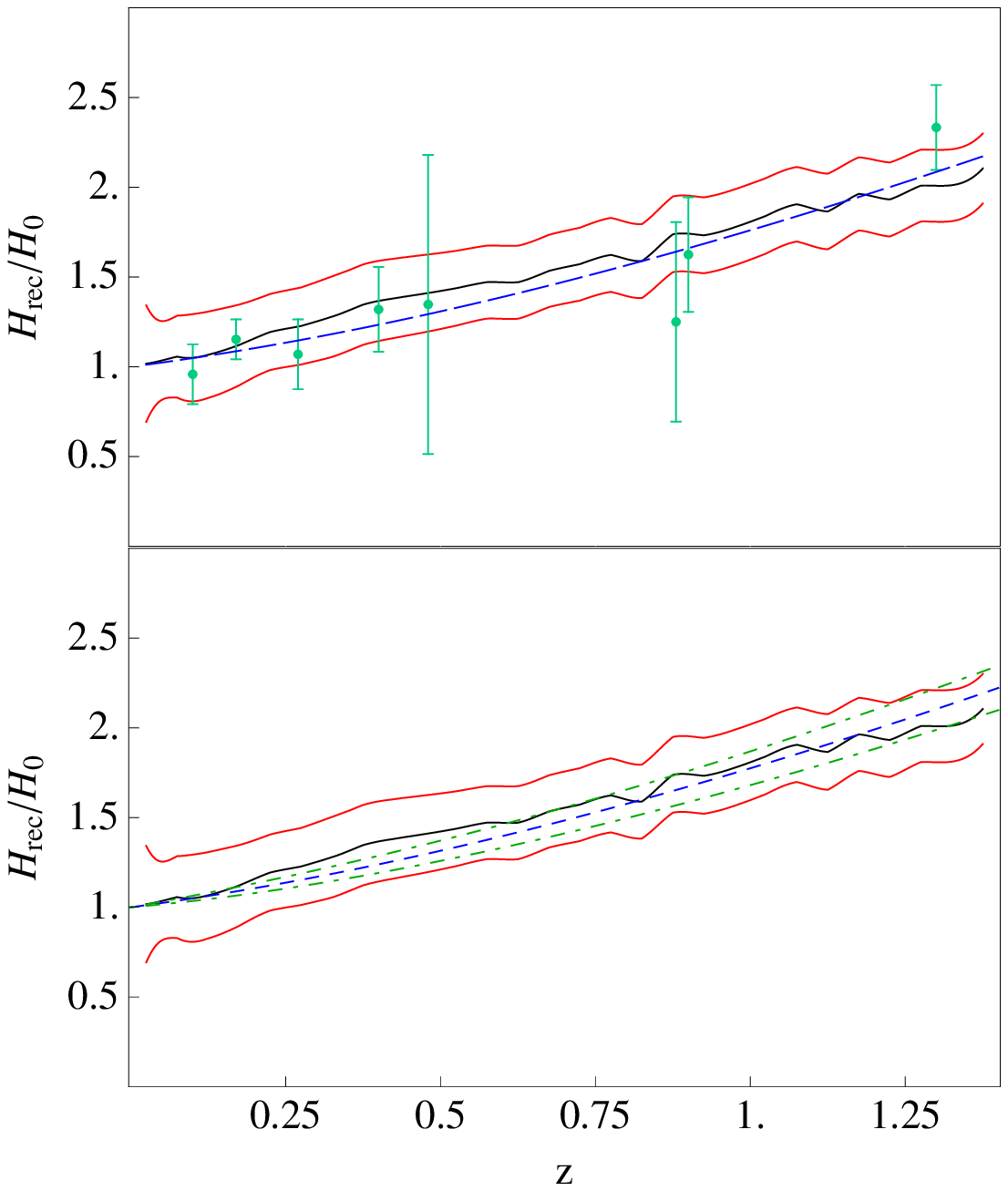}
\caption{\textbf{Top:} Reconstruction of the Huble parameter using four PCs, obtained from the real sample, superimposed on direct \emph{Hubble} parameter measurements reported by \citet{stern10} (green dots).
\textbf{Bottom:} Reconstruction of $H(z)$ with four PCs and confidence intervals of $2\sigma$ obtained from the propagation of statistical uncertainties on  $\Omega_m$ and $w$ reported by \cite{kessler09}, for MLCS2k2 results. In both panels, the blue dashed line represents $H(z)$ in a flat XCDM cosmological model with $w=-0.76$ and $\Omega_m=0.30$ (best fit results reported by \citet{kessler09}). Black curve corresponds to the best-fit reconstruction using 4 PCs  and red lines are $2\sigma$ confidence levels.} \label{fig:SDSS_REG_kes}
\end{figure}

To compare our results with other model-independent determinations of $H(z)$, we plot in the top panel of Fig. \ref{fig:SDSS_REG_kes} the reconstruction with four PCs, superimposed on measurements of $H(z)$ derived from red-envelope galaxies observations by \citet{stern10}. The error bars associated with these data points are still pretty large, but they already provide important insights into the behavior of the \emph{Hubble} parameter in the redshift range covered by the real data sample. We can see that the reconstruction encloses the fXCDM model, as well as agreeing with the red-envelope galaxy measurements. To compare our results with the predictions of a standard model-dependent procedure, we show in the bottom panel of Fig. \ref{fig:SDSS_REG_kes}, the $2\sigma$ confidence levels derived from the error propagation of statistical uncertainties in $\Omega_m$ and $w$ reported by \cite{kessler09} assuming a fXCDM model and using MLCS2k2.  We again find good agreement between the two results.

In Figs. \ref{fig:SDSS_rec6} and \ref{fig:SDSS_REG_kes}, we point out that the blue-dashed line does not represent the behavior we are trying to achieve, but is merely a representation of a model we are used to dealing with. The purpose of showing it here is to provide an idea of how far our results are from others presented in the literature, although, in our particular analysis, no assumptions about the energy content of the Universe is necessary.

The determination of what kind of physical and/or systematic effect generates patterns seen in the two lower panels of Fig. \ref{fig:SDSS_rec6} is beyond the  scope of this work. In our interpretation, these results confirm that fXCDM provides a good first-order approximation of the real behavior of $H(z)$  within the current observational errors and assumptions underlying our procedure. However, a more realistic simulation and detailed study of systematic errors are necessary in order to fully understand second-order effects.

\section{Conclusions}
\label{sec:conclusions}

We have presented an alternative procedure for extracting  cosmological parameters from type Ia supernova data. Our  analysis  is concentrated in the \emph{Hubble} parameter, although we emphasize that the same procedure can be applied to other quantities of interest. Our goal has been to be as general as possible, so we have tried to avoid parametric forms or specific cosmological models by using PCA.

Writing $H(z)$ according to equation (\ref{eq:expanso em H}) and considering type Ia supernova observations, we have shown that  it is possible to obtain analytical expressions for the Fisher matrix. We used a mock sample formed by 34 redshift bins of width $\Delta z=0.05$,  with errors calculated following the prescription proposed by \cite{kim04}. This mock sample represents a simplification of future data sets, such as the JDEM, and is not a realistic representation of current data. Our goal in using it was to check the consistentency of our procedure.

Our first attempt in reconstructing the \emph{Hubble} parameter as a linear combination of the eigenvectors of \textsf{\textbf{F}} was unsuccessful. In trying to fit high-redshift data with PCs that go asymptotically to zero, the most oscillatory modes propagate their behavior to the reconstructed $H(z)$ in the whole redshift range. As a consequence, the final result barely resembles our fiducial model.

To suppress the influence of the high-redshift behavior present in all PCs of interest, we considered the value of the \emph{Hubble} parameter at high redshift as an extra free parameter in our analysis. This simple modification provided reliable results when used with simulated and real supernova data. Beyond that, our results are corroborated with measurements of red-envelope galaxies from \citet{stern10}.

As a final remark, we emphasize that PCA provides a viable way of avoiding phenomenological parameterizations. It represents one of the few statistical methods that allow us to obtain the behavior of a chosen quantity directly from the data. It has its own assumptions, such as Gaussianity, independence of data points and in the specific case analyzed here, cosmologies that obey a FRW metric. In the final reconstruction phase, it also exhibits a bias in the upper redshift bound. On the other hand, the procedure proposed here can drastically suppress the influence of this bias. Beyond that, we show that in the context of this work, the Fisher matrix can be analytically obtained. This avoids all uncertainties related to numerical derivations of step functions and might be a good alternative to standard statistical analyses applied to cosmological data.

\begin{acknowledgements}
E.E.O.I. thanks the Brazilian agency CAPES (1313-10-0) for financial support. R.S.S.  thanks the Brazilian agencies FAPESP (2009/05176-4) and CNPq (200297/2010-4) for financial support. We also thank the anonymous referee for fruitful comments and suggestions.  This work was supported by World Premier International Research Center Initiative (WPI Initiative), MEXT, Japan.
\end{acknowledgements}

\bibliographystyle{aa}
\bibliography{ref}
\end{document}